# Europa Exploration Philosophy

A white paper submitted to

The Committee on the Planetary Science Decadal Survey (2023-2032) of

The National Academies of Sciences

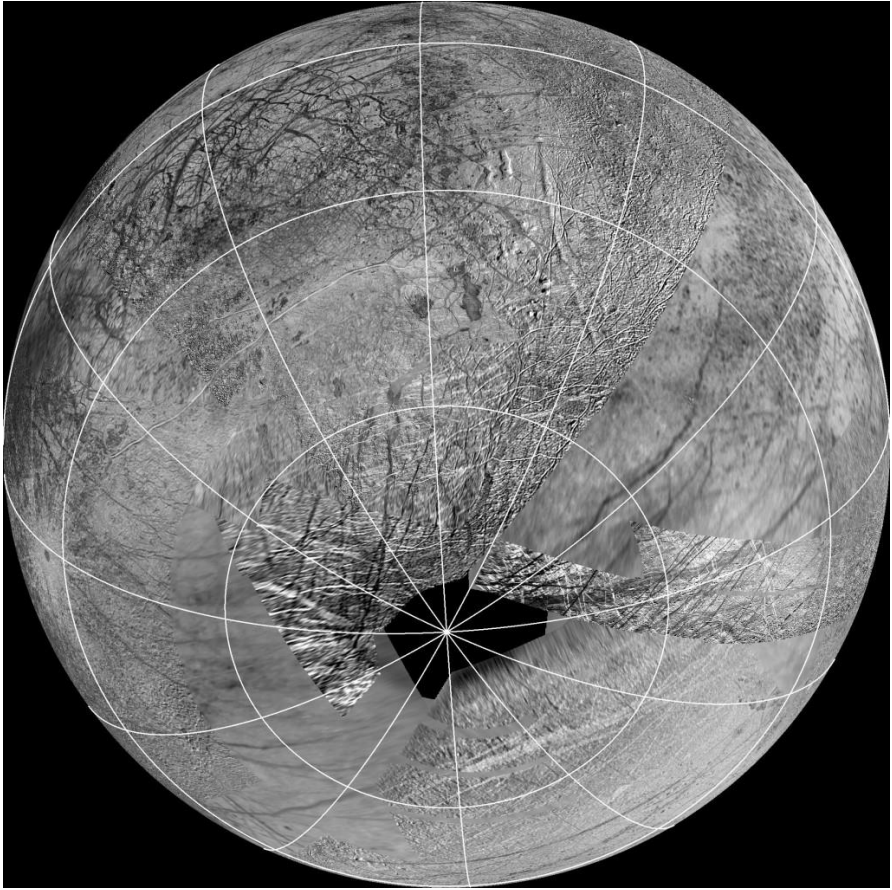

**Philip Horzempa**

**LeMoyne College; Syracuse, New York**

**September 15, 2020**

Email: horzempa45@gmail.com

**Europa Exploration Philosophy**

The exploration of Europa after the completion of the Clipper mission should be pursued by a series of low-cost scouts. These spacecraft will ascertain the nature of Europa's surface at a scale of meters to centimeters. Some will search for signs of organics and/or life. All of them will precede a large lander.

**Approach**

A Viking-class Europa lander is a high-risk, high-cost mission. NASA has pursued initial designs at the behest of Congress. A large $5 billion Europa Lander was not recommended by the most recent Decadal report and should not be recommended by the 2020 Decadal team. In its place, NASA should launch a series of small, low-cost precursor missions in the coming decade. They would determine the true nature of Europa's surface and detect any signs of life in an interior ocean. This information will allow an informed decision on whether to proceed to a Viking-class lander.

**Factors for Consideration**

If the search for life is the scientific "reason for being" for such a mission, then there are faster, better, cheaper ways of pursuing that goal than with a complex, Flagship-scale lander. There are 2 risks to proceeding with a large, costly lander. The first is our lack of knowledge of the surface conditions. The second is the assumption that there will be any organic material located in the upper meter of Europa's ice crust.

The best place to begin the search for Europan life might be several miles above the surface, in the plumes that may have been produced by an interior ocean. A scaled-down version of the Enceladus Life Finder flyby probe might be the best way to sample those plumes.

The budget line that sends precursor probes to Europa should also send the same class of small craft to Enceladus. If the goal is to find extra-Terran life in an interior ocean of an ice moon, then the best targets may be Enceladus and Miranda. Europa may be a "dry well." Investing billions of dollars solely on Europa may be a diversion from more promising water moons. Recent studies indicate that the sea floor crust on Europa may be under such enormous overburden pressure that cracks and fissures are unlikely to form. (1) That, in turn, would inhibit (or even prevent) the formation of features such as hydrothermal vents.

Europa may have an interior ocean, but it may be a biological desert. In order to make an informed decision regarding which Ice Moons to prioritize in the search for life, a set of low-cost precursor missions are necessary.

Some may assume that because a Viking-class Europa lander would use hardware (such as the MSL Skycrane) inherited from previous projects, that its cost would be reasonable. The record of 50 years of such assumptions indicates just the opposite. Examples that range from the Mars

Observer ($2 billion) to the JWST ($10 billion) abound.  Heritage hardware does not lower the cost when it is used in a new configuration.  That unique spacecraft architecture, and its associated software, must be thoroughly tested.

In the place of a complex, expensive lander, a series of simple, but capable, craft should be sent to scout the surface of Europa (and Enceladus).  The March of Progress now makes this possible.

**Landing Sites on Europa**

The nature, and hazards, of the surface of Europa are unknown.  The presence of boulders, fissures, ice spikes, and steep slopes would present dangers to the safe touchdown of any lander.

The 2012 Europa Lander Mission report concluded with this observation, "At present, there are only 15 Galileo images of Europa's surface at a resolution of 9-12 m/pixel and only 1 image at 6 m/pixel."  It further points out that "the existence of a 10-meter-square area with mild slope and limited rock distribution...is unknown."  Furthermore, "Europa appears to have very difficult landing terrain and it is conceivable that…an acceptable landing site might not be found." (2)

Smart, autonomous, guidance during descent and landing has been proposed as a method to avoid such dangers.  This terrain relative navigation (TRN) is feasible, to a degree.  It will be of no help if a smooth area is not found during descent to the surface.  A lander does not have an infinite supply of fuel that would allow it to "divert" to a safe site.

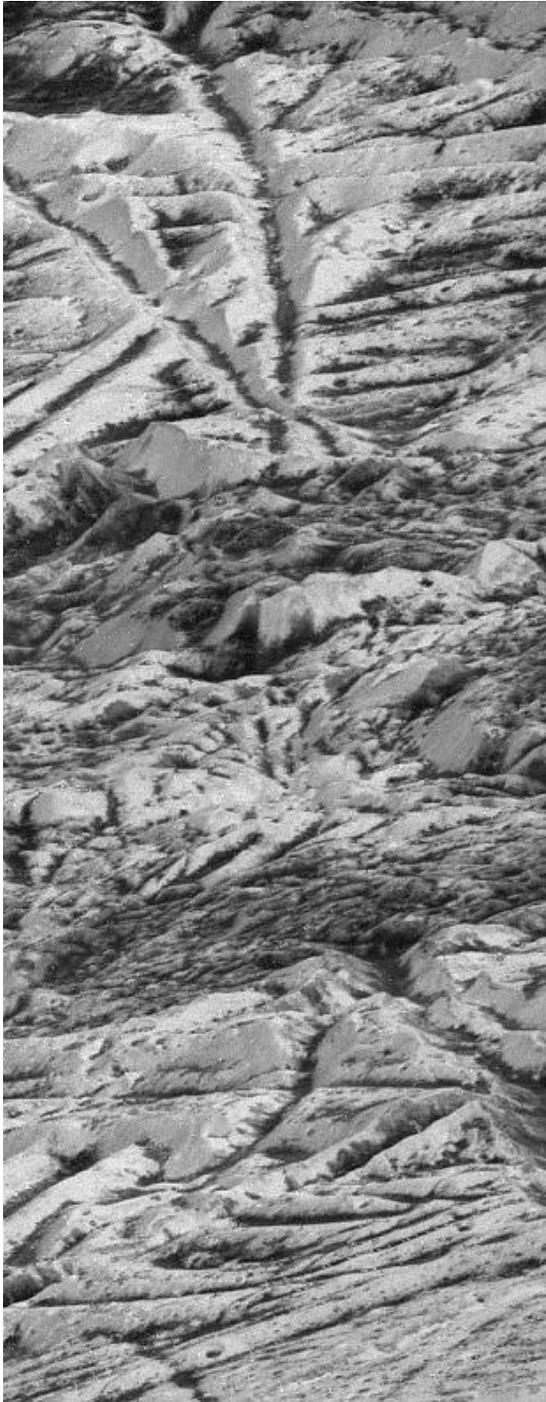
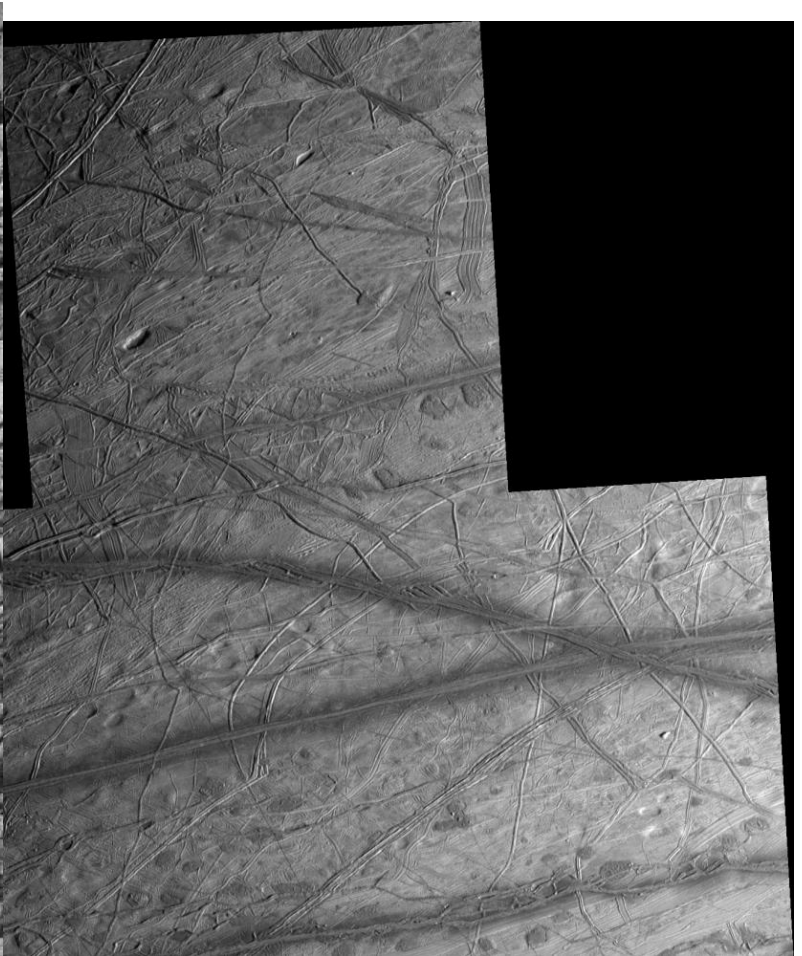

*Regional mosaic from Galileo orbit E15*

**The Surface of Europa at Close Range**

Photos from Galileo reveal little, or nothing, about the nature of Europa's surface at the scale of a lander. Obstacles that would destroy such a vehicle may be widespread. Our Moon's surface was found to include stretches of smooth terrain. One factor that produced such topography is the absence of tectonic activity. The constant rain of micrometeoroids over eons of time produced a layer of regolith that was tens of meters thick.

There are also large smooth areas on Mars. Like the Moon, it has a very low level of tectonic activity. Even though Mars' atmosphere shields it from micrometeoroids, the winds on that planet erode bedrock and produce deposits of sand.

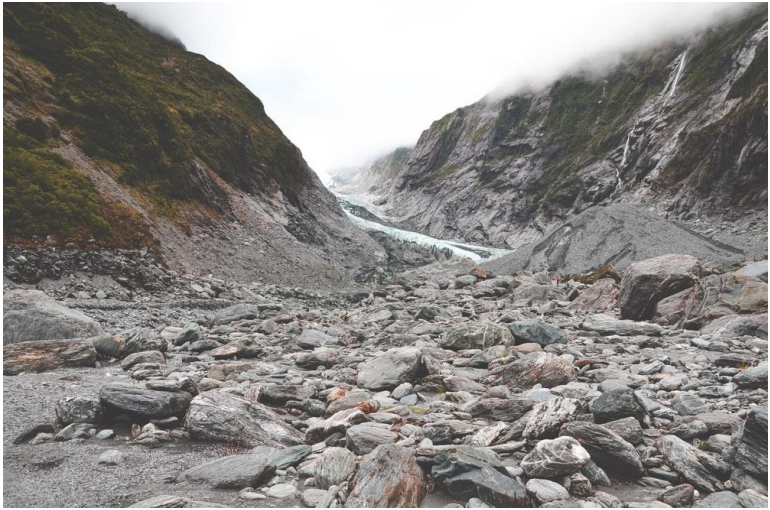
Europa has an environment unlike that of the Moon or Mars. It has undergone weathering from micrometeoroids, but it shows signs of active tectonics. The crust of Europa may be created by the freezing of liquid water. Such a phase change in the hard vacuum, and cryogenic temperatures, of Europa may produce surface features unlike any encountered by previous planetary or lunar landers. It may resemble the rift zone on Iceland, another area of recent tectonic activity. The key factor is that Europa's surface is young and has undergone very limited erosion since its formation. The "debris" created by active crustal formation and movement has had little time to wear away. Most of Europa's surface may present hazards like those found in rift zones or in the ejecta blankets of fresh craters.

**Scouts**

When faced with a similar uncertainty in the 1960s, NASA launched the Lunar Orbiter and Ranger probes to the Moon. Detailed photographs were needed to find a safe landing site for Apollo's Lunar Module. A similar approach is called for in the case of Europa, Enceladus and the other Ice Moons. Three separate classes of Scouts should be sent:

The first would be vertical-descent crash landers (**Rangers**) whose flight plan resembles that of the lunar namesake of the 1960s. They would utilize narrow-angle and context cameras to capture views of the surface to the point of impact. Modern technology would allow the high-resolution camera to produce images of features (rocks, fissures, spires) as small as 1 cm.

Side-looking cameras would capture panoramic views of the terrain during descent for further characterization of the surface.

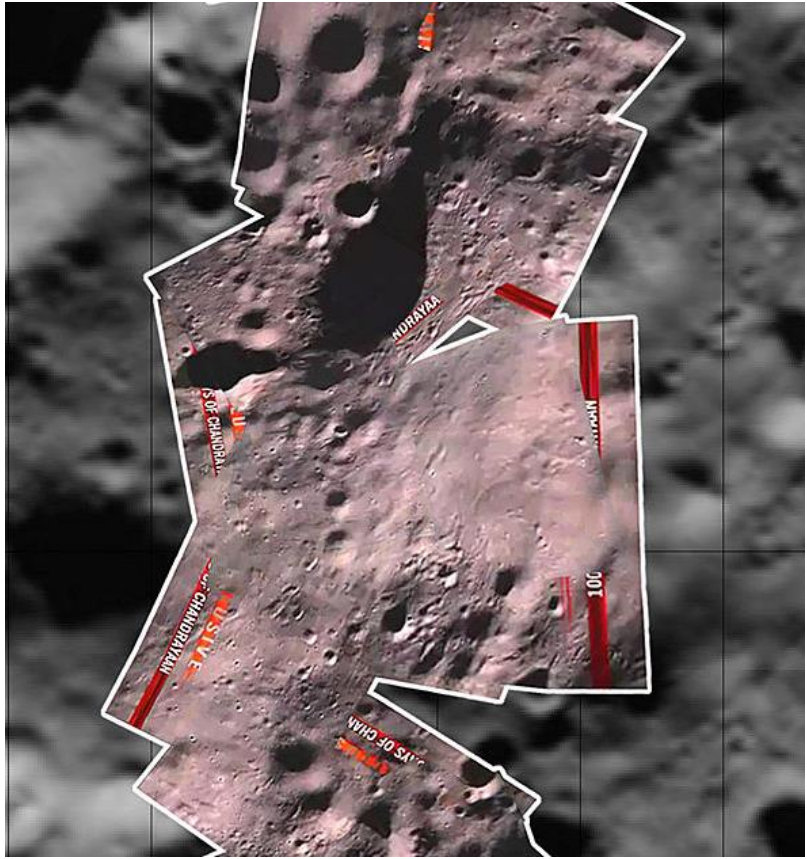
The second class would be the **Barn Stormers**, skimming as close as 100 meters above the surface. The absence of an atmosphere on Europa (and most other Ice Moons) allows for this option. They would be equipped with both nadir-pointed and horizon-oriented cameras. Their trajectory would produce a transect across 100 kilometers of terrain. Horizontal velocities of several kilometers per second limits their resolution to a degree, but images will be produced that show the presence of boulders, fissures and slopes at the scale of a future lander.

The Indian Moon Impact Probe (MIP), part of the Chandrayan-1 mission, did exactly that in 2009, as shown above.

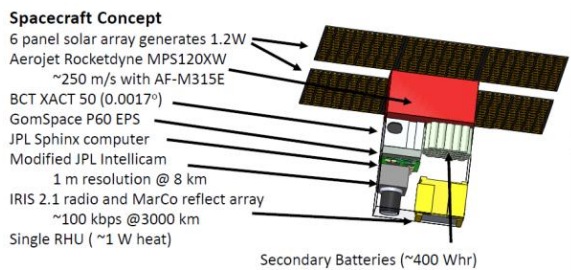

If biology is the main reason to land on Europa, then a simpler approach would be to sample the interior liquid ocean by means of **Plume Sampler** scouts. If there really are plumes generated by Europa, then the Clipper mission will detect them. Assuming that the plumes are composed of water from Europa's subsurface ocean, such a fly-through vehicle will directly analyze their composition.

A Plume Sampler will be a small sat, perhaps similar to the proposed Enceladus South Pole Imager. (3) Like the ESPI, the Europa Plume Samplers will require a relay spacecraft to downlink their data.

### Europa Scout Lander

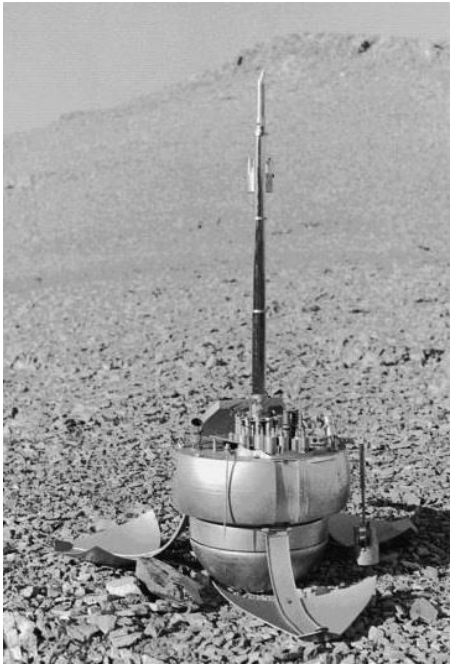

A small, rugged, simple lander could also be sent to Europa to ascertain the small-scale nature of its surface. Again, the Ranger project provides inspiration. Several of the craft were equipped with a capsule that could survive a landing at 200 mph. None of them were able to demonstrate that ability because of issues with the carrier. Such a capsule could be made to survive a non-soft landing on Europa. Equipped with several cameras, it would provide ground truth regarding the nature of the surface.

The unknown small-scale nature of Europa's surface will dictate the design of such a pathfinder lander. There are several ways to approach that challenge, but the common theme is to keep the design simple, but robust.

If a simple **Europa Scout Lander** is flown, then it could be equipped with a Laser-Induced Breakdown Spectrometer (**LIBS**) instrument. This device would produce a good deal of information on the constituents of Europa's rocks, both rocky and icy, without the need of a sampling device and sample delivery hardware. It can reveal peaks of Carbon, and other elements, that indicate the presence of organic molecules. A LIBS instrument would be capable of micron-level spatial resolution, allowing the analysis of individual particles embedded in an ice matrix. (4)

### Conclusions

All of these small/medium class probes can be flown for a fraction of the $5 billion cost of a Viking-class lander. They should be flown first in order to generate the knowledge of the surface and interior of Europa that is required before committing to a large lander.

These small missions would benefit from the presence of a relay orbiter in Jupiter space. If the Io Observer is chosen in the ongoing Discovery competition, then it would be in the right place at the right time. There is an option for the Io Observer to enter a distant orbit around Jupiter, as part of an extended mission, acting as a Jupiter system observer. With financial assistance from NASA Headquarters, the Io Observer could be equipped with a relay antenna.

The same series of low-flyby and impact Rangers should be sent to Enceladus. The absence of information on the small-scale roughness of Europa's surface applies equally to Enceladus.

**Recommendations**

  The 2020 Planetary Decadal team is asked to endorse a series of small, simple Europa scouts.

These will determine the nature of the moon's surface at the scale of a lander.  They will also sample any plumes, looking for the presence of organic chemicals.  A Viking-class lander should be pursued only if these precursor missions indicate that it is worth the cost and effort.

   The Viking-class Europa lander was inspired by the search for life.   Such a lander would cost $5 billion.  That large investment of resources in the financial aftermath of the 2020 Pandemic should be approached with caution.  If, like Viking, such a Flagship mission failed to detect Life or Organics, the repercussions for NASA's Planetary division could be severe.

  As it now stands, we know nothing about the small-scale nature of Europa's surface, or of any Ice Moon for that matter.  Before complicated Flagship landers are sent to any of these worlds, it would be prudent to send scouts ahead.

**References**

>>  1.  Paul Byrne et al (2020); "Icy Satellite Seafloors, Interiors and Habitability"; 2023-2032 Decadal Survey White Paper

>>  2.  Scott Hubbard et al; Europa Mission Review Board (2012); "Independent Technical Review Findings"; Appendix D.4.5 of the "Europa Study 2012 Report: Europa Lander Mission"

>>  3.  Andrew Blocher et al (2017); "Saturn Swarm Study: Small Probe and CubeSat Architectures to accompany New Frontiers missions at Saturn"

>>  4.  S.G. Pavlov et al (2011); "Miniature laser-induced plasma spectrometry for planetary in situ analysis – The Case for Jupiter's moon Europa"